\documentclass[11pt]{cernrep}
\usepackage{graphicx}
\usepackage{here}
\begin{document}
 \title{Jet and Dijet Rates in pA Collisions}
\author{A. Accardi$^a$ and N. Armesto$^b$}
\institute{$^a$ Institut f\"ur Theoretische Physik der Universit\"at
Heidelberg, Germany\\
$^b$ Departamento de F\'{\i}sica, Universidad de C\'ordoba, Spain
}
\maketitle

The study of jet production has become one of the precision tests of
QCD. Next-to-leading order (NLO) computations have been successfully
confronted with experimental
data. For example, jet production at the TeVatron \cite{cdfjets}
provides a very stringent test of available QCD Monte Carlo
codes and gives very
valuable information on the behaviour of the gluon distribution in the
proton at large $x$.

On the other hand, jets have never been measured in a nuclear
environment. There are prospects to measure high-$p_T$ particles at
RHIC, but only at the LHC the opportunity of detecting and studying
jets with $E_T\sim
100$ GeV will appear. Concretely, such measurements in pA collisions
would allow to settle some uncertainties like the validity of collinear
factorization \cite{facto}, the modifications of nucleon parton density
functions (pdf)
inside nuclei \cite{emc}, or the existence and amount of energy loss of fast
partons inside cold nuclear
matter \cite{quench}.
Such issues have to become well determined before any claim, based on
jet studies, on the
existence of new physics in nucleus-nucleus collisions
(hopefully QGP formation), can be considered conclusive.
Furthermore, the study of high-$p_T$ particles  and jets
can help not only to disentangle the
existence of such new state of matter but also to characterize its
properties, see e.g. \cite{tomo}.

For our computations we will use the Monte Carlo code at NLO of
\cite{nlocode,nlocode2}, adapted to include isospin effects and
modifications of nucleon pdf inside nuclei.
This code is based on the subtraction method to cancel the infrared
singularities between real and virtual contributions;
full explanations and a list of available codes, as well as detailed
discussions on theoretical uncertainties in nucleon-nucleon
collisions, can be found in
\cite{qcdyel}. The accuracy of our computations is limited by CPU time,
but it can be estimated to be:
\begin{itemize}
\item For the transverse energy distributions, 2~\% for the lowest
and 15~\% for the highest $E_T$-bins.
\item For the pseudorapidity distributions, 3~\%.
\item For the dijet distributions of the angle between the two hardest
jets, 20~\% for the least populated and 3~\% for the most populated
bins.
\end{itemize}
The results will be presented in the LHC lab frame, i.e. that in which
the experiments work. For example, for pPb collisions this means a 7 TeV proton
beam against a 2.75 TeV Pb beam, see the Section on the experimental
parameters in pA collisions.
All the energies will be given per
nucleon and, in order to compare with the pp case, all cross sections
will be presented per nucleon-nucleon pair, i.e. divided by AB.

Unless explicitly stated, we will use as nucleon pdf the MRST98 central
gluon distribution \cite{mrst98} modified inside nuclei using the EKS98
parameterizations \cite{eks98}, a factorization scale equal to the
renormalization scale $\mu=\mu_F=\mu_R=E_T/2$ (with $E_T$ the total
transverse energy of all the jets in the generated event),
and for jet reconstruction we will employ
the $k_T$-clustering algorithm \cite{ktal} with $D=1$, which is
more sound on theoretical grounds than the cone algorithm \cite{cone}.
The kinematical regions we
are going to consider are the following:
\begin{itemize}
\item $|\eta_i|<2.5$, with $\eta_i$ the pseudorapidity of the jet; this
corresponds to
the acceptance of the central part of the CMS detector.
\item $E_{Ti}>20$ GeV in the pseudorapidity distributions, with
$E_{Ti}$ the transverse energy of the jet; this will ensure the validity
of perturbative QCD.
\item $E_{T1}>20$ GeV and $E_{T2}>15$ GeV for the $\phi$-distributions,
with $E_{T1}$ ($E_{T2}$) the transverse energy of the hardest
(next-to-hardest) jet entering the CMS acceptance, and $\phi$ the angle
between these two jets.
\end{itemize}
Please note that, even in the absence of nuclear effects, the
$\eta_i$-distributions may be asymmetric with respect to $\eta_i=0$ due to
the fact that we are giving our results in the lab frame, not in the
center-of-mass frame.

Concerning the dijet distributions in the angle between the two hardest
jets, asymmetric cuts have been imposed to avoid the generation of large
logarithms in certain points
of the phase space, see \cite{nlocode2}. Also, the results near $\phi=\pi$ are
not reliable \cite{nlocode2}, as they require an all-order resummation
not available in the code we are using.

Some call for caution, which will be even more important for AB
collisions (see the Section on Jet and Dijet Rates in AB Collisions
\cite{ABcoll}), about the possible comparison of our results with the
experimental situation has to be stressed at this point.
There exist several physical effects that are not included in our
computations. For example, the so-called underlying event (i.e. the soft
particle production that coexists with the hard process) is not
considered here; it may cause difficulties with jet reconstruction and
increase the uncertainties in jet-definition algorithms. This effect
has been considered in antiproton-proton collisions
at the TeVatron \cite{under} but its estimation, which relies on our limited
knowledge of soft multiparticle production, is model dependent. In
collisions involving nuclei the situation is even worse, as our
knowledge is more limited. The only way to take this into account
would be to use available Monte Carlo simulators for the full pA event
\cite{simulators}, suitably generalized to introduce NLO QCD
calculations. 
Another process not included in our computations is multiple
hard parton collisions \cite{mhps}. They may be classified into two classes: 
\begin{itemize}
\item[a)] Disconnected collisions, i.e.
processes in which in the same event there are 
more than one independent parton-parton collisions producing 
a pair of high-$p_T$ jets each. 
This aspect has also been studied at the
TeVatron \cite{cdfmhps} for hard collisions coming from one 
nucleon-nucleon collision, but its
quantitative explanation and the extension to collisions involving
nuclei is model dependent.
Simple estimates of the influence of disconnected collisions on jet production
in pA collisions may be obtained by computing the number $\langle n \rangle$ of
nucleon-nucleon collisions involved in the production of jets with
$E_{Ti}$ greater than a given $E_{T0}$. In the Glauber model 
\cite{ccrit} one obtains: $\langle n \rangle
(b,E_{T0})=AT_A(b)\sigma(E_{T0})/\sigma_{pA}(b,E_{T0})$, with $b$ the
impact parameter, $T_A(b)$
the nuclear profile function normalized to unity, $\sigma(E_{T0})$ the
cross section for production of jets with
$E_{Ti}$ greater than $E_{T0}$ in pp collisions, and
$\sigma_{pA}(b,E_{T0})=1-[1-T_A(b)\sigma(E_{T0})]^A$. Taking
$\sigma(E_{T0})=120$, 0.3
and 0.02 $\mu$b as representative values in pPb collisions at 8.8 TeV
for $E_{T0}=20$, 100 and 200 GeV respectively (see results
in Fig. \ref{pafig5-7} below), the number
of nucleon-nucleon collisions involved turns out to be 1.0 for all
$E_{T0}$,
both for minimum bias collisions (i.e. integrating numerator and
denominator in $\sigma_{pA}(b,E_{T0})$ between $b=0$ and $\infty$)
and for central collisions (e.g. integrating between $b=0$ and 1 fm).
So, in pA collisions at LHC
energies the contribution of multiple hard scattering coming from
different nucleon-nucleon collisions seems to be negligible for the
transverse energies of the jets considered in this study.
\item[b)] Rescatterings, i.e. processes 
in which a given high-$p_T$ parton may undergo several hard collisions before
hadronizing into a jet, mimicking a single process of higher order in the
perturbative QCD expansion. Quantitative descriptions of the resulting
modification of the jet transverse energy spectrum are model
dependent. However, the effect at LHC is in general expected to be very small,
and almost negligible at the transverse energies considered in this paper
(see
the Section on Cronin effect in proton-nucleus collisions: a
survey of theoretical models \cite{roce}).
\end{itemize}
As a last comment, no centrality dependence has been studied. It is not
clear how to implement such a dependence in theoretical computations,
because not only the number of binary nucleon-nucleon collisions should
change with centrality, but also the modification of nucleon pdf
inside nuclei may be centrality-dependent.

\section{$d\sigma/dE_T$ and $d\sigma/d\eta$ for jets at large $E_T$}

\subsection{Uncertainties}

\begin{figure}[htb]
\begin{center}
\includegraphics[width=12.5cm,bbllx=0pt,bblly=30pt,bburx=560,bbury=545pt]{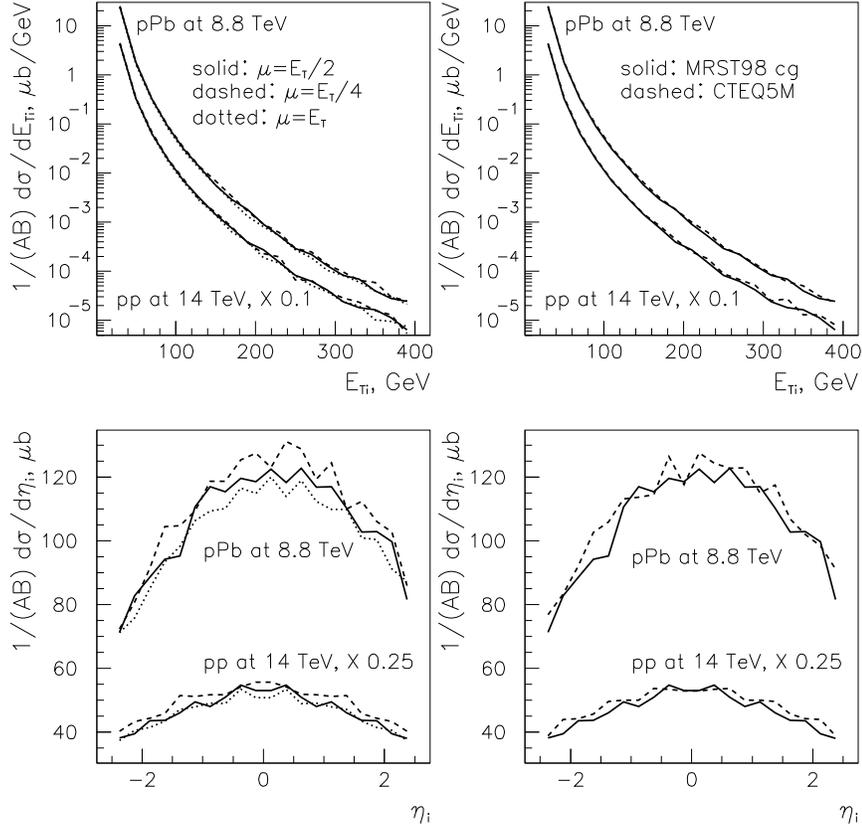}
\caption{{\it Plots on the left}:
Scale dependence of jet cross sections
($\mu=E_T/2$: solid lines; $\mu=E_T/4$: dashed
lines; $\mu=E_T$: dotted lines) versus
transverse energy of the jet (for $|\eta_i|<2.5$, upper plot) and
pseudorapidity of the jet (for $E_{Ti}> 20$ GeV, lower plot).
{\it Plots on the right}:
Nucleon pdf dependence of jet cross sections (MRST98 central gluon: solid lines;
CTEQ5M: dashed lines) versus
transverse energy of the jet (for $|\eta_i|<2.5$, upper plot) and
pseudorapidity of the jet (for $E_{Ti}> 20$ GeV, lower plot).
In each
plot, results for pPb at 8.8 TeV (upper lines) and pp at 14 TeV (lower
lines, multiplied by 0.1 in the $E_{Ti}$-plot and by 0.25 in the
$\eta_i$-plot) are shown. Unless otherwise stated default options are
used, see text.}
\label{pafig1-2}
\end{center}
\end{figure}

\begin{figure}[htb]
\begin{center}
\includegraphics[width=12.5cm,bbllx=0pt,bblly=30pt,bburx=560,bbury=545pt]{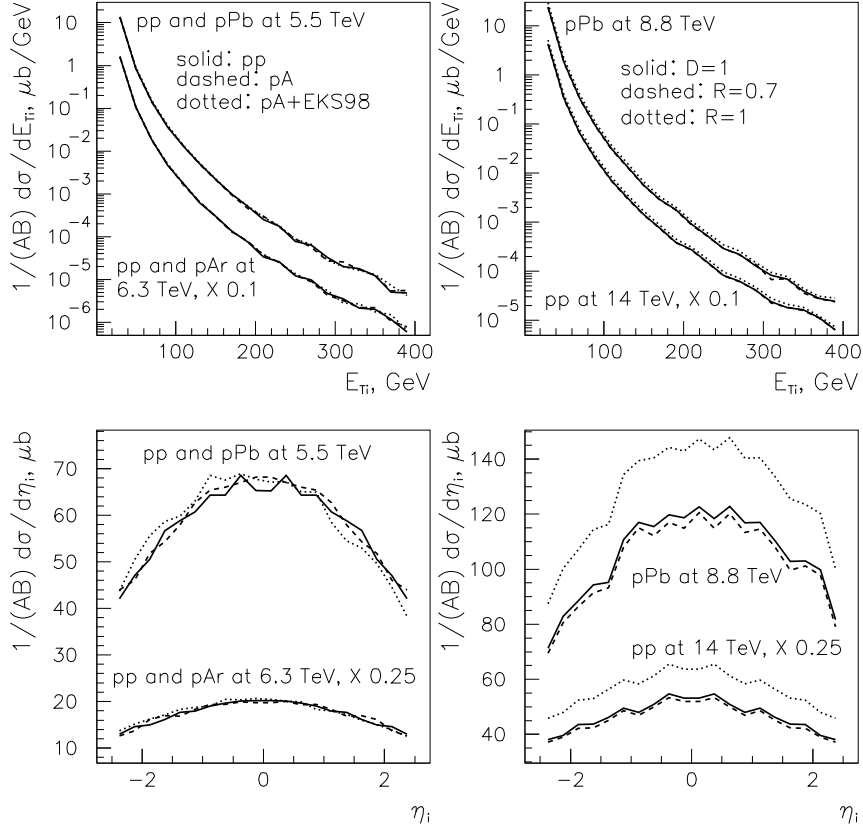}
\caption{{\it Plots on the left}:
Isospin and nuclear pdf dependence of jet cross sections (pp results:
solid lines; pA results without modification of nucleon pdf inside
nuclei: dashed lines; pA results with EKS98 modification of nucleon pdf
inside nuclei: dotted lines) versus
transverse energy of the jet (for $|\eta_i|<2.5$, upper plot) and
pseudorapidity of the jet (for $E_{Ti}> 20$ GeV, lower plot). In each
plot,
results for  pp and pPb at 5.5 TeV (upper lines) and pp and pAr at
6.3 TeV (lower
lines, multiplied by 0.1 in the $E_{Ti}$-plot and by 0.25 in the
$\eta_i$-plot) are shown.
{\it Plots on the right}: Dependence of jet cross sections
on the jet reconstruction algorithm
($k_T$-algorithm
with $D=1$: solid lines; cone algorithm with $R=0.7$: dashed lines; cone
algorithm with $R=1$: dotted lines)
versus
transverse energy of the jet (for $|\eta_i|<2.5$, upper plot) and
pseudorapidity of the jet (for $E_{Ti}> 20$ GeV, lower plot). In each
plot,
results for pPb at 8.8 TeV (upper lines) and pp at 14 TeV (lower
lines, multiplied by 0.1 in the $E_{Ti}$-plot and by 0.25 in the
$\eta_i$-plot) are shown. Unless otherwise stated default options are
used, see text.}
\label{pafig3-4}
\end{center}
\end{figure}

Four sources of theoretical uncertainties in pp and pPb collisions are
examined in Figs. \ref{pafig1-2} and \ref{pafig3-4}. First, it can be seen
that varying the choice of the scale between $E_T/4$ and $E_T$ gives
differences of order $\pm 10$~\%, the smaller scale giving the larger
results. Second, the variation due to a different choice of nucleon
pdf has been examined by using CTEQ5M as nucleon pdf \cite{cteq5},
which produces larger results
than the default choice MRST98 central gluon \cite{mrst98},
the variation being of
order 5~\%. Third, the change due to isospin effects (obtained from the
comparison of pp and pPb without any modifications of nucleon pdf
inside nuclei at the same energy per nucleon) is very small as one
would expect from the dominance of the gluon-gluon production channel;
the effect of modifications of nucleon pdf inside nuclei,
estimated by using EKS98 \cite{eks98} nuclear corrections, is also small
but produces an extra asymmetry (an excess in the region $\eta_i<0$)
in the pseudorapidity distributions for pA
collisions at reduced energy (e.g. pPb at 5.5 TeV),
which
disappears at maximum energy (e.g. pPb at 8.8 TeV), see Fig.
\ref{pafig3-4} (bottom-left) and Fig. \ref{pafig5-7} (bottom-right),
due to the the
asymmetry of momentum of projectile and target in the LHC lab frame.
Finally, the choice of jet-finding algorithm produces some differences:
the cone algorithm \cite{cone} with $R=0.7$ gives results slightly
smaller ($\sim 1\div2$~\%) than the $k_T$-clustering algorithm
\cite{ktal} with $D=1$, while cone with $R=1$ gives results $\sim 15$~\%
larger than our default choice.

\begin{figure}[htb]
\begin{center}
\includegraphics[width=12.5cm,bbllx=0pt,bblly=30pt,bburx=560,bbury=545pt]{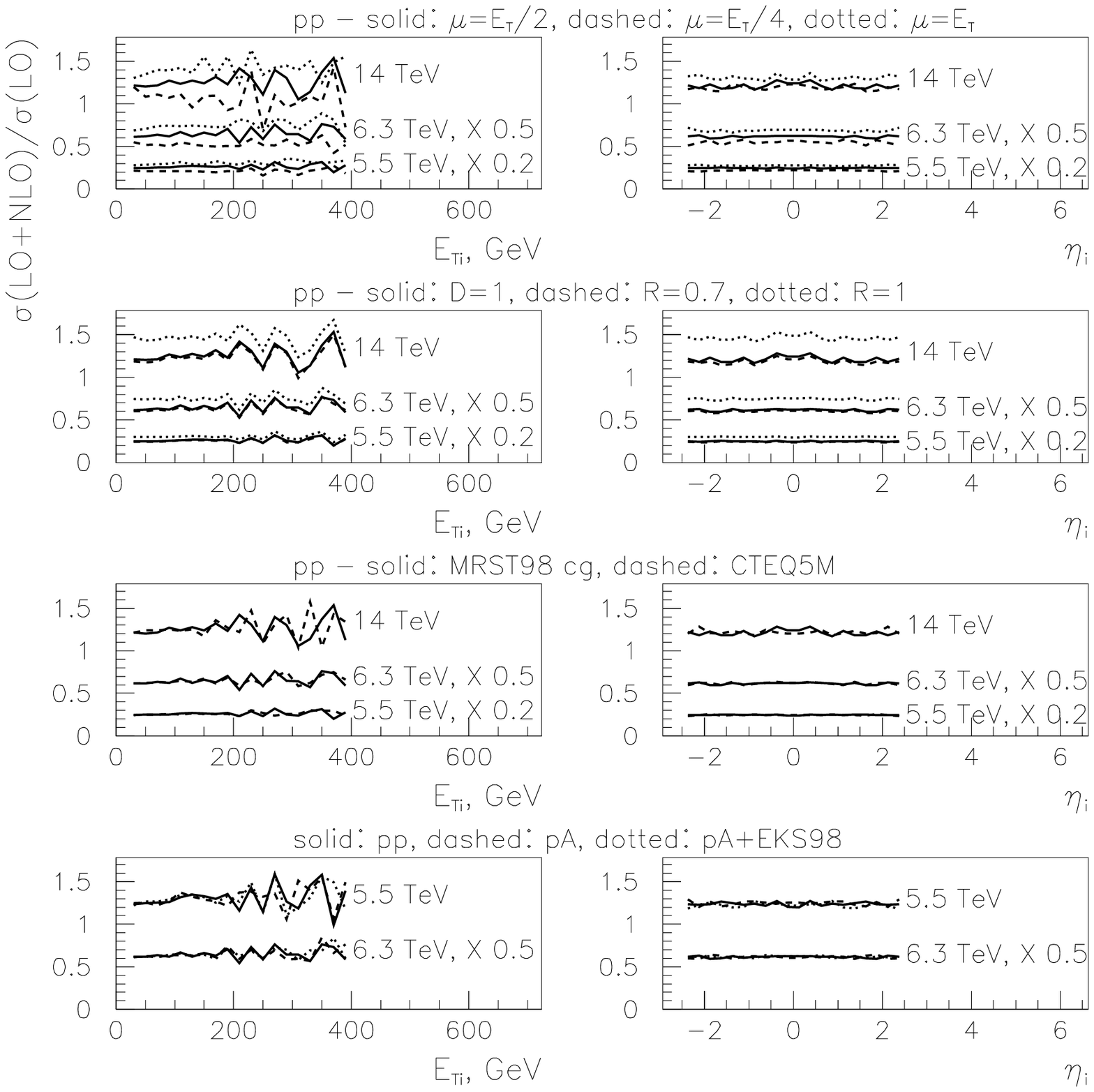}
\caption{Ratio of cross section computed at NLO over
those computed at
LO versus the transverse energy of the jet (for $|\eta_i|<2.5$, plots on the
left) and the pseudorapidity of the jet (for $E_{Ti}> 20$ GeV, plots
on the right).
From top to bottom:
scale dependence
($\mu=E_T/2$: solid lines; $\mu=E_T/4$: dashed
lines; $\mu=E_T$: dotted lines); dependence
on the jet reconstruction algorithm
($k_T$-algorithm
with $D=1$: solid lines; cone algorithm with $R=0.7$: dashed lines; cone
algorithm with $R=1$: dotted lines); nucleon pdf dependence (MRST98
central gluon: solid lines; CTEQ5M: dashed lines); and isospin and nuclear
pdf dependence (pp results:
solid lines; pA results without modification of nucleon pdf inside
nuclei: dashed lines; pA results with EKS98 modification of nucleon pdf
inside nuclei: dotted lines).
{\it Six upper plots}: results for pp collisions at 14, 6.3
(multiplied by 0.5) and 5.5 (multiplied by 0.2) TeV are shown. {\it
Plots at the bottom}:
results for pp and pAr collisions
at 6.3 TeV (multiplied
by 0.5), and for pp and pPb collisions at 5.5 TeV, are given.
Unless otherwise stated default options are
used, see text.}
\label{kfact}
\end{center}
\end{figure}

As a last comment, let us discuss the ratio of cross sections evaluated
at NLO over those computed at leading order (LO), the so-called
$K$-factor. In Fig. \ref{kfact} this ratio has been examined for
different collisions and energies, and varying the different sources of
uncertainties. For our default option the ratio results quite
constant with the transverse energy or pseudorapidity of the jet and
turns out to be around $1.2$ for the energies examined. The dependence on
the renormalization/factorization scale results in a ratio $\sim 1.35$ for
$\mu=E_T$ and $\sim 1.15$ for $\mu=E_T/4$. The cone algorithm with
$R=0.7$ gives results very similar to those obtained with the
$k_T$-clustering algorithm
with $D=1$, while cone with $R=1$ produces a ratio $\sim
1.45$ (note that at LO the choice of jet-finding algorithm has no
influence on the results).
Finally, neither a variation of nucleon pdf, nor isospin effects
and modifications of nucleon pdf inside nuclei, have a sizeable effect.

\subsection{Results}

\begin{figure}[htb]
\begin{center}
\includegraphics[width=16.0cm,bbllx=0pt,bblly=30pt,bburx=545,bbury=535pt]{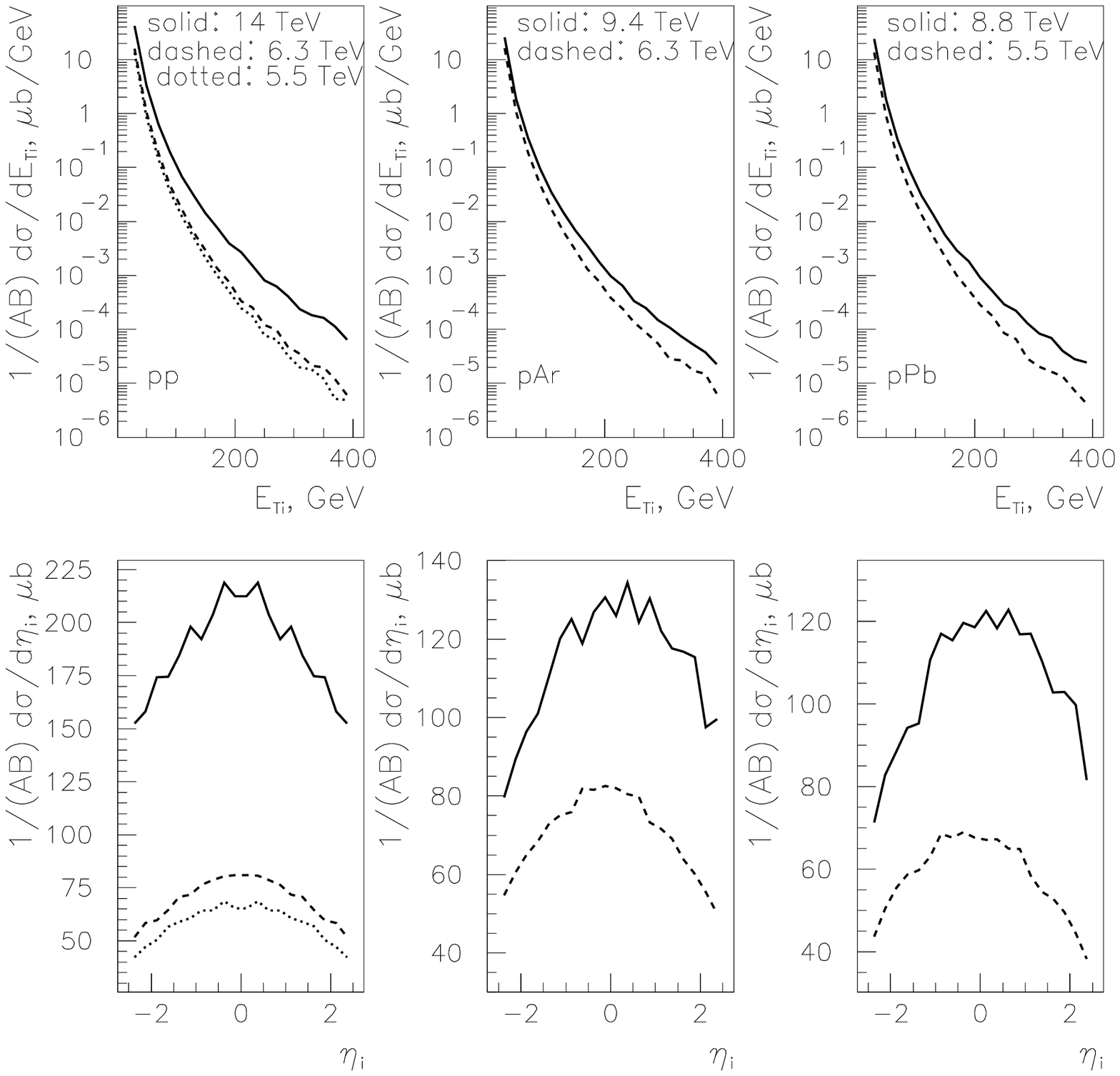}
\caption{Energy dependence of jet cross sections
versus
transverse energy of the jet (for $|\eta_i|<2.5$, upper plots) and
pseudorapidity of the jet (for $E_{Ti}> 20$ GeV, lower plots). {\it Plots
on the left}: pp
collisions at 14 (solid lines), 6.3 (dashed lines)
and 5.5 TeV (dotted lines). {\it Plots in the middle}:
pAr
collisions at 9.4 (solid lines)
and 6.3 TeV (dashed lines). {\it Plots on the right}: pPb
collisions at 8.8 (solid lines)
and 5.5 TeV (dashed lines). Default
options are
used, see text.}
\label{pafig5-7}
\end{center}
\end{figure}

In Table 1 the number of expected events with at least one jet with a
given $E_{Ti}> 20$ GeV and $|\eta_i|< 2.5$ (or with two jets $(1,2)$ with
$E_{T1}>20$ GeV, $E_{T2}>15$ GeV
and $|\eta_{1,2}|<2.5$
for the dijet $\phi$-distributions),
per $\mu$b and pair of
colliding nucleons is shown for different collisions and possible
luminosities.
From this Table and using the Figures it is possible to know the
number of expected events with a given kinematical variable. For
example, examining the solid line in Fig.~\ref{pafig5-7} (upper-left)
one can expect, within the pseudorapidity region we have
considered, the following numbers of jets in pp collisions at 14 TeV
for a luminosity of $10^{34}$ cm$^{-2}$s$^{-1}$: $10^{10}$
jets with $E_{Ti}\sim 70$ GeV (corresponding to a cross 
section of 1 $\mu$b/(AB)), and $10^6$ jets with $E_{Ti}\sim
380$ GeV (corresponding to a cross section of $10^{-4}$ $\mu$b/(AB)).

Looking at the results given in
Fig. \ref{pafig5-7}, it becomes evident that the study of samples of
$\sim 10^3$ jets should be feasible, from a theoretical point of view,
up to a transverse energy $E_{Ti}\sim 325$ GeV with a run of 1 month
at the considered luminosity: indeed, looking at 
Table 1, $10^3$ jets for pPb at 8.8 TeV would 
correspond to a cross section of $4.8\cdot 10^{-5}$ $\mu$b/(AB), which
in Fig. \ref{pafig5-7} (upper-right) cuts the solid curve at
$E_{Ti}\sim 325$ GeV. 

\begin{table}[tbh]
\begin{center}
Table 1: Luminosities and expected number of events
with at least one jet with a
given $E_{Ti}> 20$ GeV and $|\eta_i|< 2.5$ (or with two jets $(1,2)$
with
$E_{T1}>20$ GeV, $E_{T2}>15$ GeV
and $|\eta_{1,2}|<2.5$
for the dijet $\phi$-distributions),
per $\mu$b/(AB)
in one
month ($10^6$ s), for
different collisions.
\vskip0.2cm
\begin{tabular}{|c|c|c|c|}
\hline
Collision & $E_{cm}$ per nucleon (TeV) & $\cal{L}$
(cm$^{-2}$s$^{-1}$) & Number of events per month per $\mu$b/(AB) \\
\hline
pp & 14 & $10^{34}$ & $10^{10}$ \\
\hline
pp & 14 & $3\cdot 10^{30}$ & $3\cdot 10^6$ \\
\hline
pAr & 9.4 & $4\cdot 10^{30}$ & $1.6\cdot 10^8$ \\
\hline
pAr & 9.4 & $1.2\cdot 10^{29}$ & $4.8\cdot 10^6$ \\
\hline
pPb & 8.8 & $10^{29}$ & $2.1\cdot 10^7$ \\
\hline
\end{tabular}
\end{center}
\end{table}

Some conclusions can be obtained at this point: First, as the
influence of modifications of nucleon pdf inside nuclei is not large,
an extensive systematic study of the A-dependence of these cross sections
does not seem to be the
most urgent need. Second, noting the asymmetry
in the pseudorapidity distributions in pPb collisions at the reduced
energy of 5.5 TeV (and to a lesser extent in pAr at 6.3 TeV) when EKS98
corrections are implemented, a pA run at
reduced proton energies might offer a possibility to test the influence of the
modification of nucleon pdf inside nuclei; a study of
$E_{Ti}$-distributions in different $\eta_i$-slices going from the
backward (A) to the forward (p) region would scan different regions in
$x$ of the ratio of the gluon distribution inside the nucleus over that
in a free nucleon, which is badly under control with the presently
available experimental information.
Nevertheless, this would require a detailed comparison with pp results at the
same energy in order to disentangle possible biases (e.g. detector
effects);
in case these pp results were not available experimentally,
the uncertainty due to the scale
dependence
(which in our computations at reduced energies is as large as the
asymmetry) in the extrapolation from 14 TeV to the energy of the pA
collision should be reduced as much as possible for this effect to be
useful.

\section{High-$E_T$ dijet momentum imbalance}

Dijet distributions of the angle between the two hardest
jets offer a possibility to
test the perturbative expansion in QCD. At lowest order (LO)
in collinear factorization \cite{facto} the two jets are produced
back-to-back, so any deviation from a simple peak at $\phi=\pi$ in our
results is a signal of NLO corrections. But please remember that
the results near $\phi=\pi$ are
not reliable \cite{nlocode2}, as there NLO corrections become negative
and larger than LO, so this region requires an all-order resummation.

\subsection{Uncertainties}

\begin{figure}[htb]
\begin{center}
\includegraphics[width=12.5cm,bbllx=0pt,bblly=30pt,bburx=560,bbury=545pt]{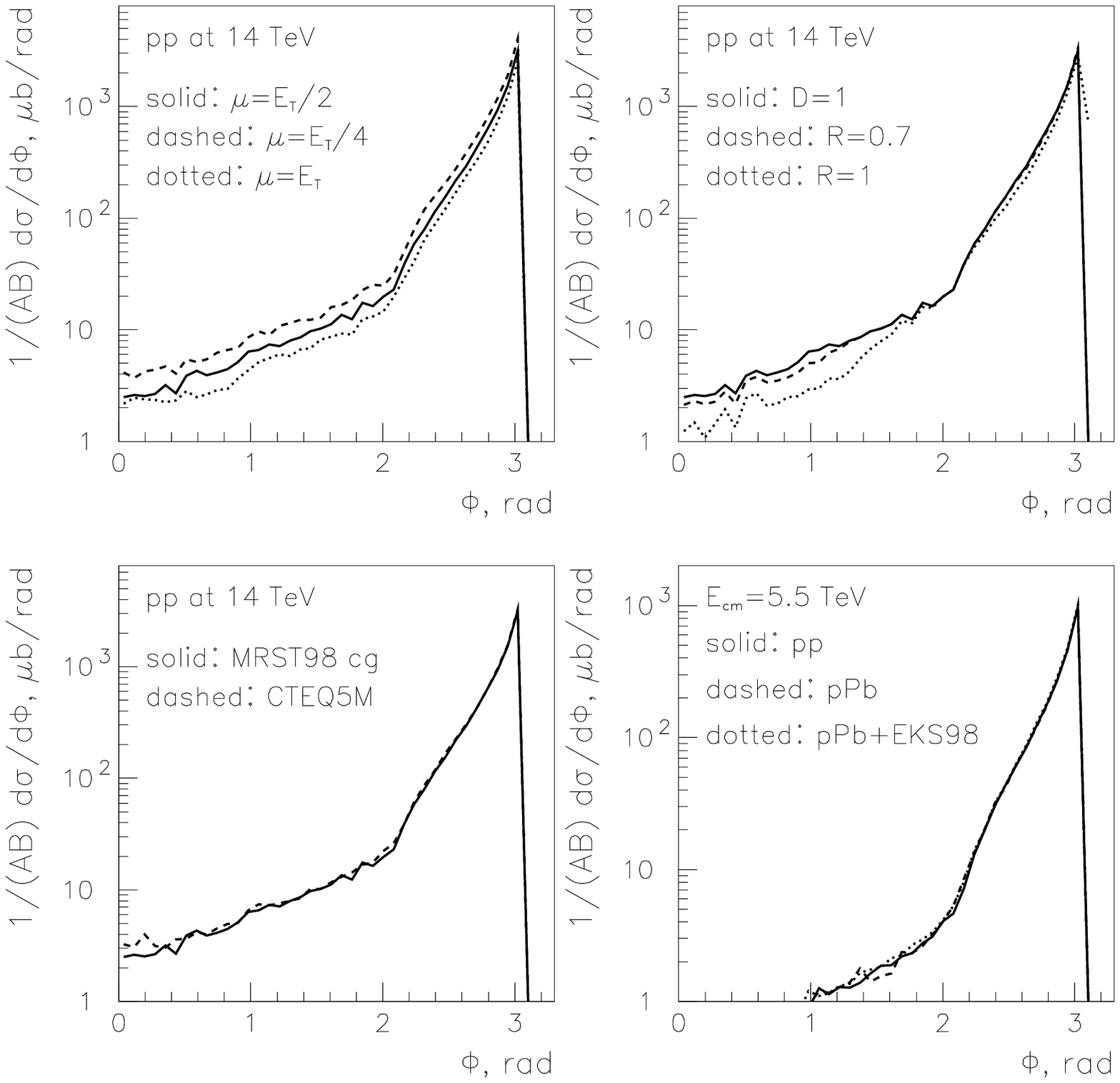}
\caption{Uncertainties in dijet cross sections
versus angle between the two hardest jets for $E_{T1}>20$ GeV,
$E_{T2}>15$ GeV and $|\eta_1|,|\eta_2|<2.5$. {\it Upper left plot}: scale
dependence in pp collisions at 14 TeV with the same line convention as
in Fig. \protect{\ref{pafig1-2}}left. {\it Upper right plot}:
dependence on the jet
reconstruction algorithm in pp collisions at 14 TeV with the same line
convention as
in Fig. \protect{\ref{pafig3-4}}right. {\it Lower left plot}:
nucleon pdf dependence
in pp collisions at 14 TeV with the same line
convention as
in Fig. \protect{\ref{pafig1-2}}right. {\it Lower right plot}:
isospin and nuclear
pdf dependence in pp and pPb collisions at 5.5 TeV with the same line
convention as
in Fig. \protect{\ref{pafig3-4}}left. Unless otherwise stated default
options are
used, see text.}
\label{pafig8}
\end{center}
\end{figure}

In Fig. \ref{pafig8} the same uncertainties examined for transverse
energy and pseudorapidity distributions are studied in the dijet cross
sections
versus angle between the two hardest jets. First, the smaller the scale,
the largest the results, so the results for
$\mu=E_T$ become up to $\sim 50$~\% smaller than for $\mu=E_T/4$. Second, the
choice of nucleon pdf seems to have a small effect (less than 10~\%),
as we have
tested by using CTEQ5M instead of MRST98 central gluon. Third, the
variation due to isospin effects (obtained from the
comparison of pp and pPb without any modifications of nucleon pdf
inside nuclei at the same energy per nucleon) is again negligible,
and the effect of modifications of nucleon pdf inside nuclei,
estimated by using EKS98 \cite{eks98} nuclear corrections, is also
small ($\sim 3$~\%). Finally, the choice of jet-finding algorithm produces
some differences which can be of order 20~\% for our three choices of
cone with $R=1$, cone with $R=0.7$ and the default $k_T$-clustering
algorithm with
$D=1$.

\subsection{Results}

\begin{figure}[htb]
\begin{center}
\includegraphics[width=12.5cm,bbllx=0pt,bblly=30pt,bburx=560,bbury=545pt]{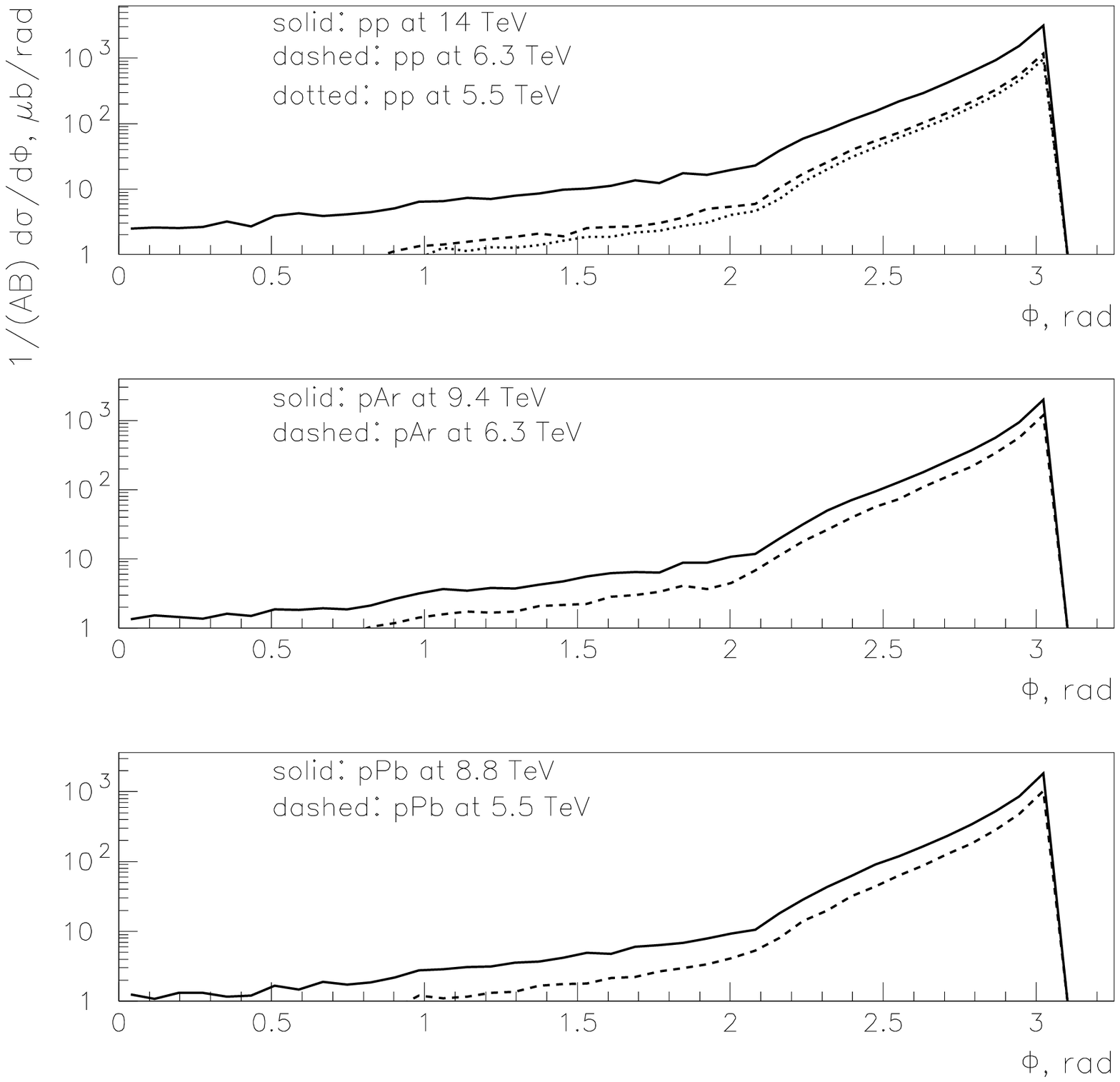}
\caption{Energy dependence of dijet cross sections
versus angle between the two hardest jets for $E_{T1}>20$ GeV,
$E_{T2}>15$ GeV and $|\eta_1|,|\eta_2|<2.5$. {\it Upper plot}: results in pp
collisions with the same line convention as in Fig.
\protect{\ref{pafig5-7}}left. {\it Medium plot}: results in pAr collisions
with the
same line convention as in Fig.
\protect{\ref{pafig5-7}}middle. {\it Lower plot}: results in pPb collisions
with the
same line convention as in Fig.
\protect{\ref{pafig5-7}}right. Default options
are
used, see text.}
\label{pafig9}
\end{center}
\end{figure}

Results for pp and pA collisions at different energies are presented in
Fig. \ref{pafig9}. From this Figure and Table 1,
the dijet momentum imbalance should
be clearly measurable, provided jet reconstruction is possible in the
nuclear environment and no other physics contribution, like the underlying
event or multiple hard parton scattering, interferes, spoiling the
comparison with experimental data of the
theoretical predictions we present.
Again, only extensive studies using Monte
Carlo simulators including full event reconstruction
will be able to clarify whether such measurements
are feasible or not, see the Section on Jet Detection at CMS.

\vskip 1cm
\noindent

\section*{ACKNOWLEDGEMENTS}
N. A. thanks CERN Theory Division, Departamento de F\'{\i}sica de
Part\'{\i}culas at Universidade de Santiago de Compostela, Department of
Physics at University of Jyv\"askyl\"a, Helsinki Institute of
Physics and Physics Department at BNL, for
kind hospitality during stays in which parts of this work have been
developed; he also acknowledges financial
support by CICYT of Spain under contract
AEN99-0589-C02 and by Universidad de C\'ordoba.
A. A. thanks the Max-Planck-Institut f\"ur Kernphysik of Heidelberg for
permission to use its computing facilities; his work is partially funded by
the European Commission IHP program under contract HPRN-CT-2000-00130.

\end{document}